\documentclass[12pt,preprint]{aastex}

\newcommand{\um}{$\mu$m}

\newcommand{\Msun}{M$_{\odot}$}
\newcommand{\Mvirial}{M$_{V}$}
\newcommand{\Lsun}{L$_{\odot}$}

\newcommand{\tco}{\mbox{$^{13}${\rmfamily CO}{\,(3--2)}}}
\newcommand{\tcont}{\mbox{$^{13}${\rmfamily CO}}}

\newcommand{\ceo}{\mbox{\rmfamily C}$^{18}${\rmfamily O}{\,(3--2)}}

\newcommand{\so}{\mbox{\rmfamily SO}{\,(7$_{8}$--6$_{7}$)}}
\newcommand{\iso}{\mbox{$^{34}${\rmfamily SO}{\,(8$_{9}$--7$_{8}$)}}}

\newcommand{\hcn}{\mbox{\rmfamily HCN}{\,(4--3)}}

\newcommand{\kms}{km~s$^{-1}$}
\newcommand{\cs}{\mbox{CS~(3--2)}}
\newcommand{\sio}{\mbox{SiO~(2--1)}}

\newcommand{\Spitzer}{{\it Spitzer}}
\newcommand{\cms}{${\rm cm}^{-2}$}
\newcommand{\cmc}{${\rm cm}^{-3}$}
\newcommand{\hii}{\mbox{$\mathrm{H\,{\scriptstyle {II}}}$}}

\newcommand{\irdcfortythree}{G034.43+00.24~MM1}
\newcommand{\irdcthirty}{G028.53$-$00.25~MM1}
\newcommand{\irdcfortythreenomm}{G034.43+00.24}
\newcommand{\irdcthirtynomm}{G028.53$-$00.25}
\newcommand{\vlsr}{V$_{LSR}$}

\newcommand{\nthp}{N$_{2}$H$^{+}$~(1--0)}
\newcommand{\htcop}{H$^{13}$CO$^{+}$~(1--0)}
\newcommand{\hcop}{HCO$^{+}$~(1--0)}
\newcommand{\nthpnt}{N$_{2}$H$^{+}$}
\newcommand{\htcopnt}{H$^{13}$CO$^{+}$}
\newcommand{\hcopnt}{HCO$^{+}$}
\newcommand{\chtchtcn}{CH$_{3}$CH$_{2}$CN}
\newcommand{\ammonia}{NH$_{3}$}
\newcommand{\tastar}{T$^{*}_{A}$}

\slugcomment{}

\shorttitle{A tale of two cores}
\shortauthors{Rathborne et al.}

\begin{document}
\title{SMA observations of Infrared Dark Clouds: A tale of two cores}
\author{J. M. Rathborne}
\affil{Harvard-Smithsonian Center for Astrophysics, Mail Stop 42, 60 Garden Street, Cambridge, MA 02138, USA; jrathborne@cfa.harvard.edu}
\author{J. M. Jackson}
\affil{Institute for Astrophysical Research, Boston University, Boston, MA 02215; jackson@bu.edu}
\author{Q. Zhang}
\affil{Harvard-Smithsonian Center for Astrophysics, Mail Stop 42, 60 Garden Street, Cambridge, MA 02138, USA; qzhang@cfa.harvard.edu}
\author{R. Simon}
\affil{I.Physikalisches Institut, Universit\"at zu K\"oln, 50937 K\"oln, Germany; simonr@ph1.uni-koeln.de} 
\begin{abstract}
We present high-angular resolution sub-millimeter continuum images and
molecular line spectra obtained with the Submillimeter Array toward
two massive cores that lie within Infrared Dark Clouds; one actively
star-forming (G034.43+00.24~MM1) and the other more quiescent
(G028.53$-$00.25~MM1). The high-angular resolution sub-millimeter
continuum image of G034.43+00.24~MM1 reveals a compact ($\sim$
0.03 pc) and massive ($\sim$ 29\,\Msun) structure while the molecular line
spectrum shows emission from numerous complex molecules. Such a rich
molecular line spectrum from a compact region indicates that
G034.43+00.24~MM1 contains a hot molecular core, an early stage in the
formation of a high-mass protostar. Moreover, the velocity structure
of its $^{13}${\rmfamily CO}{\,(3--2)} emission indicates that this B0
protostar may be surrounded by a rotating circumstellar envelope. In
contrast, the sub-millimeter continuum image of G028.53$-$00.25~MM1
reveals three compact ($\lesssim$ 0.06 pc), massive (9--21\,\Msun)
condensations but with no lines detected in its spectrum. We suggest
that the core G028.53$-$00.25~MM1 is in a very early stage in the
high-mass star-formation process; its size and mass are sufficient to
form at least one high-mass star, yet it shows no signs of localized
heating. Because the combination of high velocity line wings with a
large IR--mm bolometric luminosity ($\sim$~10$^{2}$\,\Lsun) indicates
that this core has already begun to form accreting protostars, we
speculate that the condensations may be in the early phase of
accretion and may eventually become high-mass protostars.  We,
therefore, have found the possible existence of two high-mass
star-forming cores; one in a very early phase of star-formation and
one in the later hot core phase.  Together the properties of these two
cores support the idea that the earliest stages of high-mass
star-formation occur within IRDCs.
\end{abstract}
\keywords{dust, extinction--stars:formation--ISM:clouds--infrared:ISM--radio lines:ISM}

\section{Introduction}

Emerging evidence suggests that the very earliest stages of high-mass
star and cluster formation occur within cold, dense molecular clumps
called infrared dark clouds (IRDCs;
\citealp{Simon-msxgrs,Rathborne06,Rathborne07}). Because these
molecular clumps have very high column densities ($\sim
10^{23}$--$10^{25}$ cm$^{-2}$) and low temperatures ($<25$ K), they
have predominantly been identified via their absorption of the
background Galactic emission at mid-IR wavelengths (e.g.,
\citealp{Egan98,Carey98,Carey00,Simon-catalog}).  Because IRDCs are
the densest clumps in molecular clouds \citep{Simon-msxgrs} which are
now undergoing the process of fragmentation and condensation, IRDCs
are important laboratories to study the pristine, undisturbed physical
conditions of cluster-forming clouds before they are shredded apart by
stellar winds and radiation.

Millimeter/sub-millimeter continuum studies of IRDCs
\citep{Lis94,Carey00,Garay04,Ormel05,Beuther05,Rathborne05,Rathborne06} 
reveal that IRDCs contain many compact cores. These cores have typical
sizes of $<$~0.5~pc and masses of $\sim$~120\,\Msun\,
\citep{Rathborne06}.  While most of these cores contain little evidence
for active star formation, some are associated with bright 24\,\um\,
emission, broad molecular line emission, shocked gas, and maser
emission, indicating that they are actively forming stars (e.g.,
\citealp{Rathborne05,Wang06,Chambers08}). Indeed, a number of low- and
intermediate-mass protostars \citep{Carey00,Redman03} in addition to
high-mass protostars \citep{Beuther05,Rathborne05,Pillai06,Wang06}
have been identified within IRDCs.

Hot Molecular Cores (HMCs) are associated with the early stages of
high-mass star formation. They correspond to the stage immediately
after a cold dense starless core has formed. During the later stages
in their evolution, HMCs are often also associated with methanol
masers and ultra-compact \hii\, regions. Numerous examples of HMCs
have been found throughout the Galaxy (e.g.,
\citealp{Garay99,Kurtz00,Churchwell02}), including one within an IRDC
\citep{Rathborne07}. HMCs are compact ($<$0.1
pc), dense ($\sim$10$^{5}-10^{8}$\,\cmc), and massive
($\sim$10$^{2}$\,\Msun;\citealp{Garay99,Kurtz00,Churchwell02}). Due to
the internal heating from the central high-mass protostar
(T$\sim$100\,K), they are typically very luminous ($>$10$^3$\,\Lsun)
and show strong emission from complex molecules (e.g.,
\citealp{Kurtz00}). Accretion disks are also inferred in the later
stage of the HMC phase directly \citep{Zhang05a,Cesaroni07} or
indirectly \citep{Kurtz00,Beuther02,Zhang01,Zhang05b} by the presence of
molecular outflows and maser emission.  The existence of HMCs in IRDCs
establishes a possible link between IRDCs and the early stages of
high-mass star formation \citep{Rathborne07}.

To understand the earliest stages in high-mass star formation one
needs to identify and study the so-called `high-mass starless cores,'
the immediate precursors to high-mass protostars, HMCs, and
ultra-compact \hii\, regions. High-mass star formation is rare and
occurs rapidly; thus, the identification of high-mass starless cores
is difficult. Because IRDCs have sizes, masses, and densities similar
to cluster-forming molecular clumps but are considerably colder, we
suggest that they are the precursors to clusters and their dense, cold
cores the precursors to stars. Thus, the high-mass cores within IRDCs
that contain no evidence for star formation are good candidates for
the elusive `high-mass starless cores.'

Here we present interferometric sub-millimeter continuum and molecular
line observations obtained with the Submillimeter Array toward two
cores within IRDCs; one showing evidence for active high-mass star
formation (\irdcfortythree), the other showing none (\irdcthirty).
Both cores are at distances d $>$ 3.7\,kpc and have sizes R $<$0.8 pc,
gas masses M $>$400\,\Msun, and bolometric luminosities L
$>$10$^{2}$\,\Lsun\, (Table~\ref{core-properties};
\citealp{Simon-msxgrs,Rathborne06,Rathborne08}). We speculate that the `active'
core is in a later evolutionary stage than the more quiescent core.
The new interferometric data reveal distinct properties for these
cores and suggest that \irdcfortythree\, is a more evolved high-mass
protostar in the HMC phase and that \irdcthirty\, may be in a very
early `starless' core phase of high-mass star-formation.

\section{Observations and Data Reduction}

\subsection{Interferometric continuum and spectral line data}

Interferometric observations of \irdcfortythree\, and \irdcthirty\,
were carried out with the Submillimeter Array\footnote{The
Submillimeter Array is a joint project between the Smithsonian
Astrophysical Observatory and the Academia Sinica Institute of
Astronomy and Astrophysics, and is funded by the Smithsonian
Institution and the Academia Sinica.} (SMA;
\citealp{Ho04}) on 2006 September 11 with seven antennas in the
compact configuration.  The projected baselines of the visibility data
range from 15m to 70m.  The double-sideband receivers cover a total of
4 GHz bandwidth.  The digital correlator was configured to cover 329.1
to 331.1 GHz for the lower sideband and 339.1 to 341.1 GHz for the
upper sideband with a uniform channel spacing of 0.8125 MHz
($\sim$~0.7\,\kms) across the entire band. This setup resulted in
simultaneous observations of the \tco, \ceo, \so, and \iso\,
transitions.

The 336 GHz zenith opacities, measured with the NRAO tipping
radiometer located at the Caltech Submillimeter Observatory, were
$\tau$ $\sim$ 0.4 (scaled from the 225 GHz measurement via $\tau_{336
GHz}$ = 2.8 $ \times \tau_{225 GHz}$). The measured double-sideband
system temperatures corrected to the top of the atmosphere were
between 200 and 800 K. 

The primary beam of the SMA at these frequencies is $\sim 37''$. The
phase centers for the observations are listed in
Table~\ref{core-properties}.  We used the quasars 1751$+$096 and
1743$-$038 to calibrate time dependent gains, and 3C454.3 to remove
the gain variations across the passband.  The flux scale was
referenced to Uranus.  The visibility data were calibrated with the
IDL superset MIR package developed for the Owens Valley Radio
Observatory (OVRO) Interferometer.  The absolute flux level is
accurate to about 25\%. After the calibration in MIR, the visibility
data were exported to the MIRIAD format for further processing and
imaging.  The continuum is constructed from the line free channels in
the visibility domain, and is further self-calibrated using the clean
components of the image as input models. The gain solutions from the
continuum self calibration is then applied to the spectral line data.
The rms noise in the naturally weighted maps is $\sim$ 12 and $\sim$ 4
mJy beam$^{-1}$ in the continuum (for \irdcfortythree\, and
\irdcthirty\, respectively). The noise in the continuum map for 
\irdcfortythree\, is higher because fewer line-free channels were available.
For both cores the rms noise is 120 mJy~beam$^{-1}$ ($\sim$ 2.9 K) per
1.0 \kms\, channel in the line data.  The final synthesized beam was
$\sim$ 2.3\arcsec\,$\times$ 2.1\arcsec\, which corresponds to a
physical size of $\sim$ 0.04 pc ($\sim$ 8,000~AU) and $\sim$ 0.06 pc
($\sim$ 12,600~AU) for \irdcfortythree\, and \irdcthirty\,
respectively.

\subsection{Single dish spectra}

To help with the interpretation of the SMA data, we have also included
here lower-angular resolution spectra.  These data were obtained as
part of a large molecular line survey of cores within IRDCs conducted
using the 30\,m Institute de Radioastronomie Millimetrique (IRAM)
telescope and the 15\,m\, James Clerk Maxwell Telescope (JCMT).  The
molecular lines of CS (3--2), \nthp, SiO (2--1), \htcop, and \hcop\,
were obtained with IRAM in 2004 November, while the HCN (4--3), \tco,
and \ceo\, spectra were obtained with the JCMT in 2004 April. The IRAM
spectra have a typical beam size of $\sim$ 26\arcsec, an rms noise
of 0.1 K channel$^{-1}$, and a velocity resolution of 0.07\,\kms. The
JCMT spectra have a typical beam size of $\sim$ 14\arcsec, an rms 
noise of 0.1 K channel$^{-1}$, and a velocity resolution of 1.1\,\kms.

In both cases, chopper wheel calibration, pointing, and focus checks
were performed regularly. All spectra were obtained in
position-switched mode with a suitable nearby emission-free position
used as a reference (typically $\sim$5\arcmin\, from the core).  The
data were reduced using standard methods in the software packages
GILDAS and SPECX.

\section{Results}

\subsection{Low-angular resolution molecular line emission}

\begin{figure}
\centering
\includegraphics[width=0.27\textwidth,clip=true]{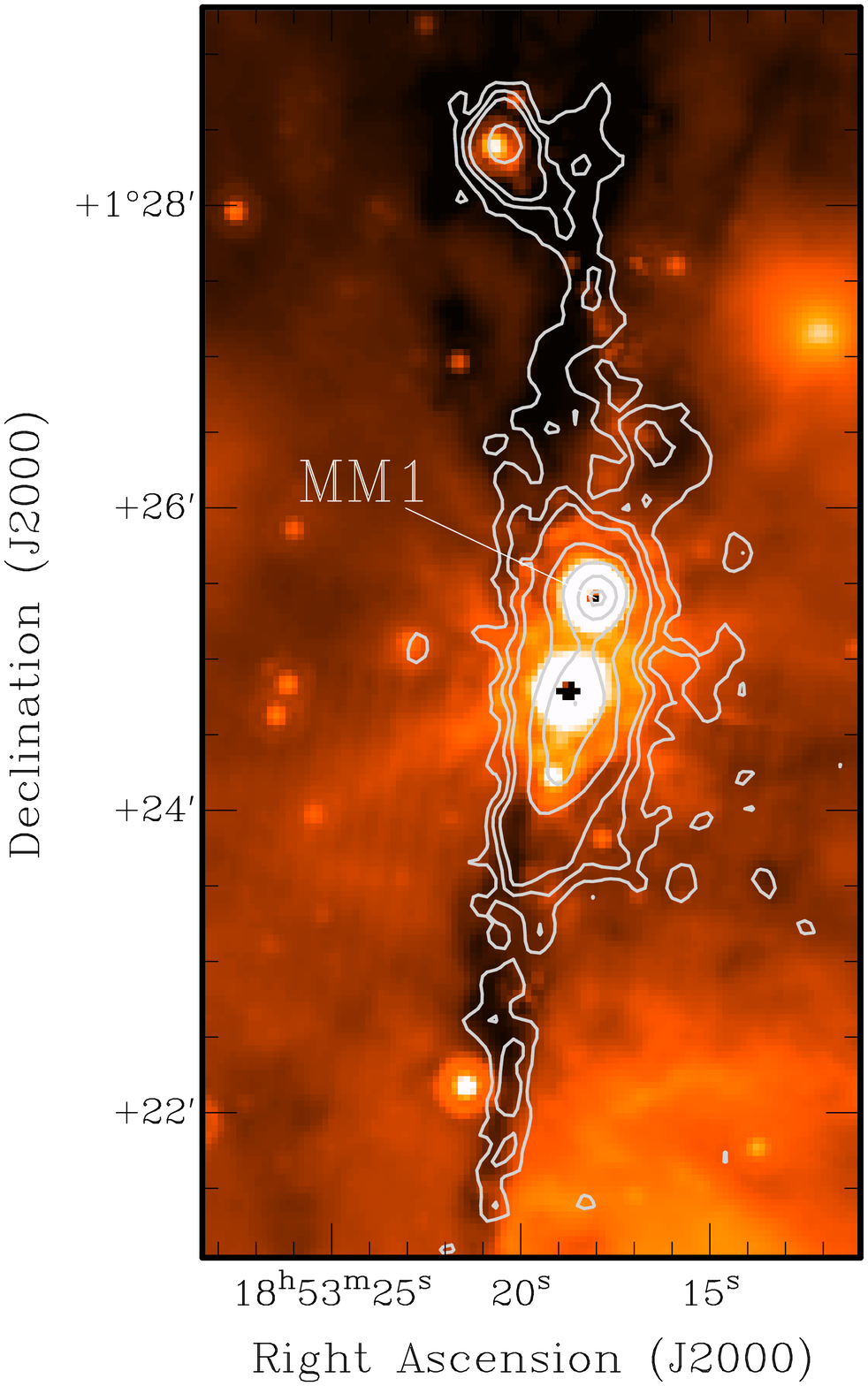}
\includegraphics[width=0.55\textwidth]{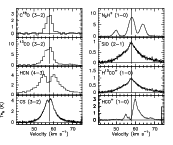}
\caption{\label{irdc-43} \irdcfortythreenomm. {\it Left}: 
 \Spitzer\, 24\,\um\, image with contours of the IRAM 30\,m 1.2~mm
 continuum emission (\citealp{Rathborne06}; contour levels are 60, 90,
 120, 240, 480, 1200, and 2200 mJy beam$^{-1}$; color scale is
 logarithmic from 30 MJy sr$^{-1}$ [black] to 200 MJy sr$^{-1}$
 [white]). Note that while the IRDC remains dark at 24\,\um\, the core
 MM1 is associated with bright 24\,\um\, emission. {\it Right}: IRAM
 and JCMT molecular line emission toward the core MM1. The solid
 vertical line marks the central velocity of the core, as traced by
 the optically thin \ceo\, emission. Note the broad line widths in the
 high-density tracers, \cs\, and \hcn, the shocked gas as traced by
 the bright \sio\, emission and the high velocity line wings in
 \hcop\, and \htcop. Table~\ref{lines} summarizes the results of
 fits to these spectra.}
\end{figure}

\begin{figure}
\centering
\includegraphics[width=0.34\textwidth,clip=true]{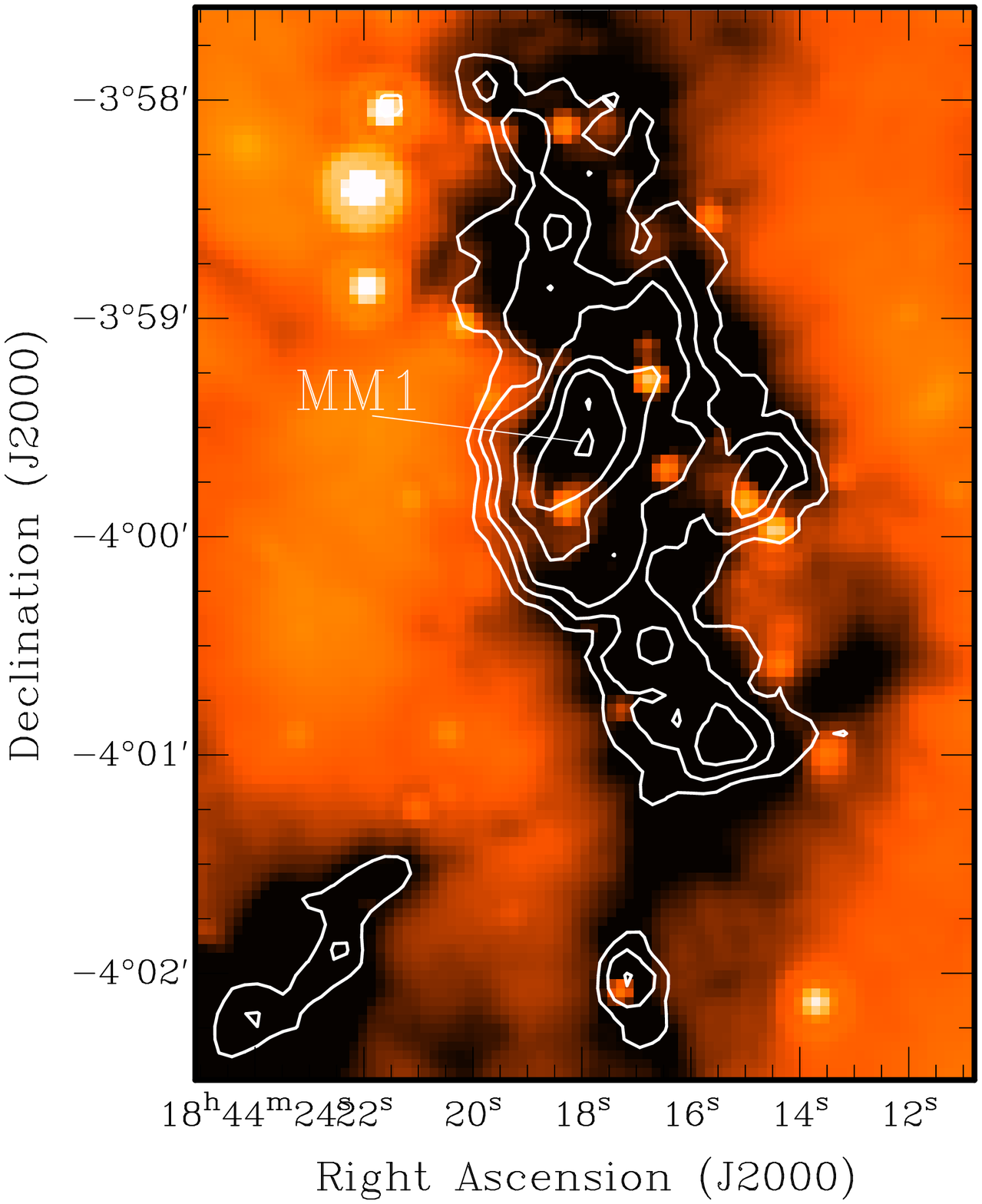}
\includegraphics[width=0.52\textwidth]{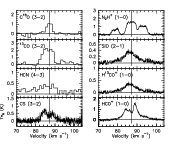}
\caption{\label{irdc-30}  \irdcthirtynomm. {\it Left}: \Spitzer\, 
 24\,\um\, image with contours of the IRAM 30\,m 1.2~mm continuum
 emission (\citealp{Rathborne06}; contour levels are 60, 90, 120, 180,
 240, and 320 mJy beam$^{-1}$; color scale is logarithmic with 50 MJy
 sr$^{-1}$ [black] to 100 MJy sr$^{-1}$ [white]). Note that the IRDC
 and, in particular, the core MM1 remain dark at 24\,\um. {\it Right}:
 IRAM and JCMT molecular line emission toward the core MM1. The solid
 vertical line marks the central velocity of the core, as traced by
 the optically thin \ceo\, emission. Note the saturated \tco\, and
 \nthp\, line emission and the faint emission from \cs\, and \hcn. The
 presence of broad line wings in the \sio, \hcop, and \htcop, spectra
 suggests that some star-formation activity is occurring within this
 core. Table~\ref{lines} summarizes the results of fits to these
 spectra.}
\end{figure}

Figures~\ref{irdc-43} and \ref{irdc-30} show the \Spitzer\, 24\,\um\,
images of the two IRDCs overlaid with the low-angular resolution
(11\arcsec) IRAM 30~m 1.2~mm continuum emission
\citep{Rathborne06}. Note that while both IRDCs remain dark at
24\,\um, the core \irdcfortythree\, shows bright emission at 24\,\um. 

Also included in these figures are the single dish low-angular
resolution molecular line spectra obtained toward the cores. In both
cases we see broad, saturated, and self-absorbed lines. In particular,
toward \irdcfortythree\, we see broad line widths in the high-density
tracers, \cs\, and \hcn, shocked gas as traced by the bright \sio\,
emission, and high velocity line wings in
\hcop\, and \htcop. Toward \irdcthirty\, we see saturated \tco\, and
\nthp\, line emission, faint emission from \cs\, and \hcn\, with broad
line wings in the \sio, \hcop, and \htcop\, spectra.

Table~\ref{lines} gives a summary of Gaussian fits to each of these
spectra: peak \tastar, central \vlsr, and FWHM line width, $\Delta
V$. Because many of the lines show non-Gaussian profiles, the listed
integrated intensity ($I$) was calculated by summing the emission over
a suitable velocity range ($\Delta V$ + 5\,\kms), rather than simply
calculating the area under the Gaussian.  For the \nthp\, spectra,
however, we do not fit a single Gaussian profile, but instead fit the
hyperfine components simultaneously. The derived opacity for the main
hyperfine components are 0.10 for \irdcfortythree\, and 0.15 for
\irdcthirty. In these cases, the integrated intensity was calculated
over the velocity range to include the emission from all hyperfine
components.

\subsection{High-angular resolution continuum images}

\begin{figure*}[ht]
\centering
\includegraphics[angle=-90,width=0.49\textwidth]{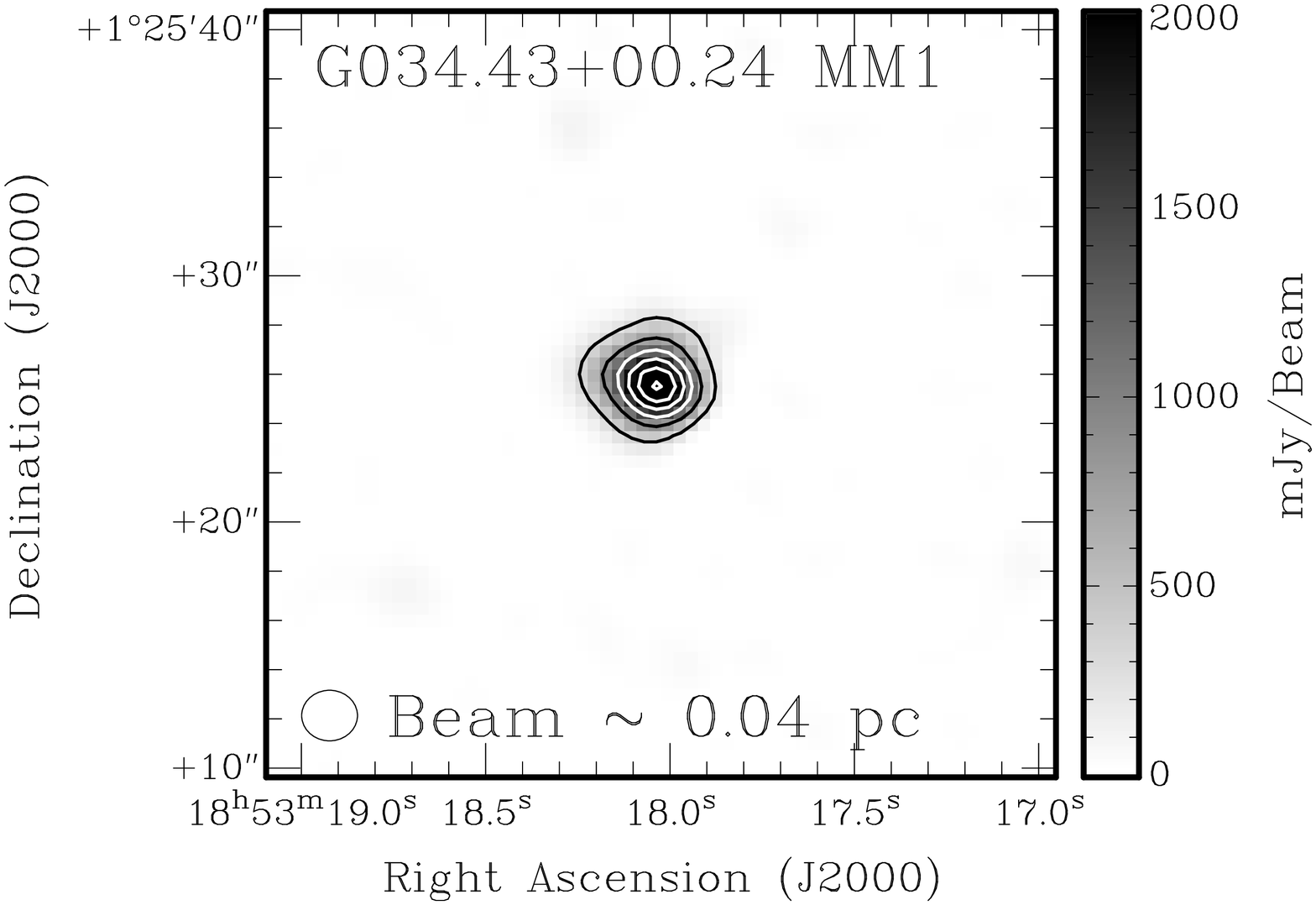}
\includegraphics[angle=-90,width=0.49\textwidth]{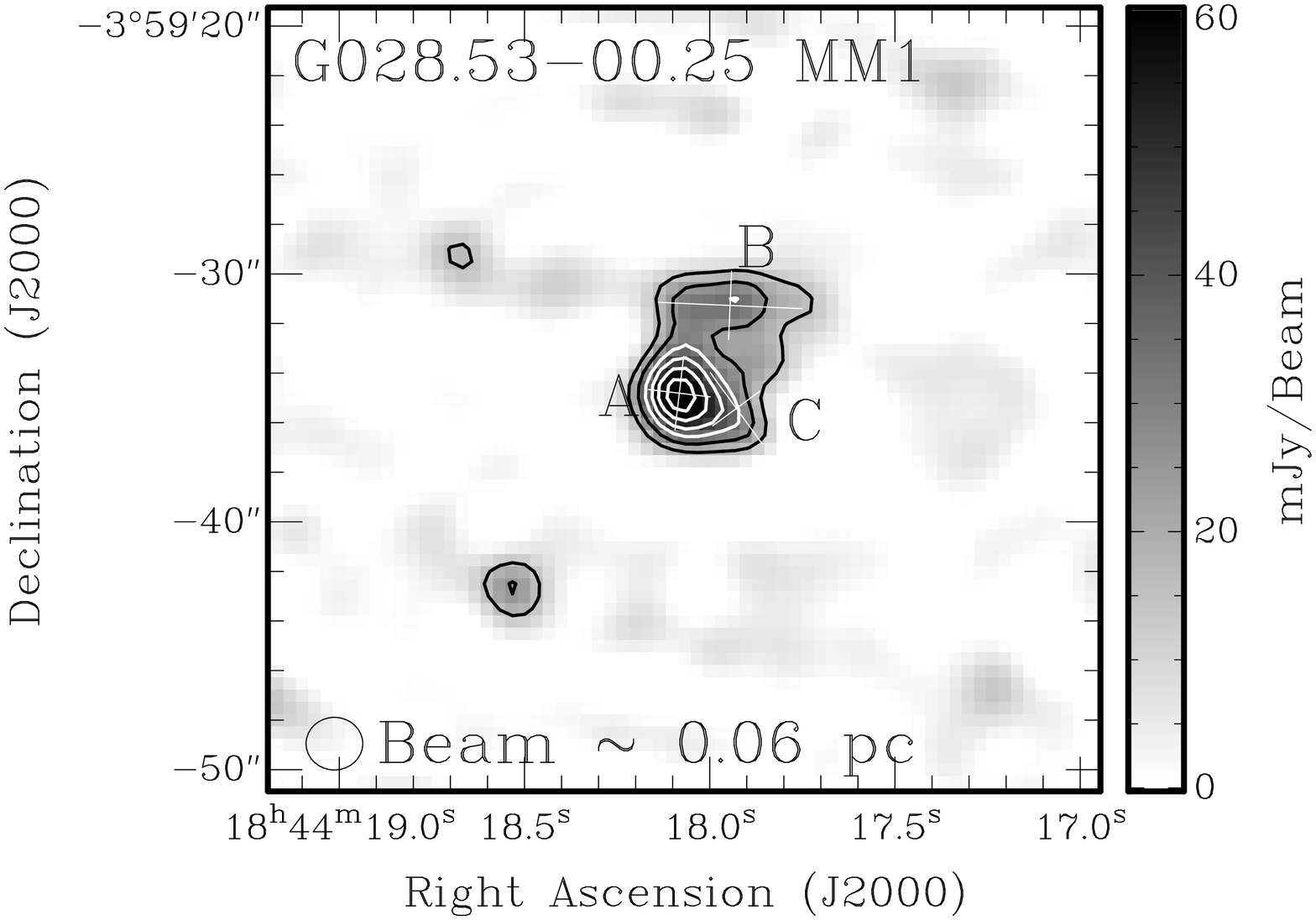}
\caption{\label{continuum-images} SMA continuum images toward the two cores 
  (the images cover $\sim$ 30\arcsec\, around the core's center).  The
  contour levels for \irdcfortythree\, (left) are 0.3 to 3.0 in steps
  of 0.5~Jy~beam$^{-1}$ and for \irdcthirty\, (right) are 0.015 to
  0.07 in steps of 0.01~Jy~beam$^{-1}$. In both cases the lowest two
  contour levels are in black for clarity.  Note that
  \irdcfortythree\, contains a single, compact ($\sim$ 0.03 pc),
  massive (24\,\Msun) emission feature, while \irdcthirty\, appears
  resolved into three individual condensations.  These condensations
  are marked with a cross and have sizes of $\sim$ 0.06 pc and masses
  of 8--19\,\Msun. Table~\ref{condensations} list their properties.}
\end{figure*}

Despite the similar sizes, masses, and bolometric luminosities (see
Table~\ref{core-properties}) derived for these cores from the
low-angular resolution continuum images, the high-angular resolution
sub-millimeter continuum images show significant differences
(Fig.~\ref{continuum-images}). Toward \irdcfortythree\, the emission
remains unresolved, while toward \irdcthirty\, the emission is
resolved into three individual condensations\footnote{In this paper we
use the term `clump' to refer to IRDCs. These large ($\sim$ 5\,pc),
massive ($\sim$ 10$^{3}$\,\Msun), dense ($\sim$ 10$^{3}$\cmc)
molecular structures are found within a Giant Molecular Cloud (GMC)
and will likely give rise to a cluster of stars. Within these
`clumps', the many compact substructures are referred to as `cores'
and have typical sizes of $\sim$ 0.5\,pc and masses of $\sim$
10$^{2}$\,\Msun. At high-angular resolution, these `cores' can often be
further resolved into multiple substructures. We refer to the small
continuum substructures identified by the SMA, with sizes $\sim$
0.05\,pc, as `condensations'. The `condensations' presumably trace the
material that will give rise to the individual stars.}.  In both cases
the emission is coincident with the peak in the lower-angular
resolution millimeter continuum images (shown as contours on
Figs.~\ref{irdc-43} and \ref{irdc-30}). Table~\ref{condensations}
lists the positions, deconvolved sizes, and peak fluxes which were
determined from Gaussian fits to the emission within the high-angular
resolution images.

\subsection{High-angular resolution molecular line emission}

The high-angular resolution molecular line data toward these two cores
are also very different. Toward \irdcfortythree\, the spectrum
contains many molecular line emission features, but, in contrast, the
spectrum toward \irdcthirty\, contains no significant molecular line
emission whatsoever.

\begin{figure*}[ht]
\centering
\includegraphics[angle=-90,width=0.99\textwidth]{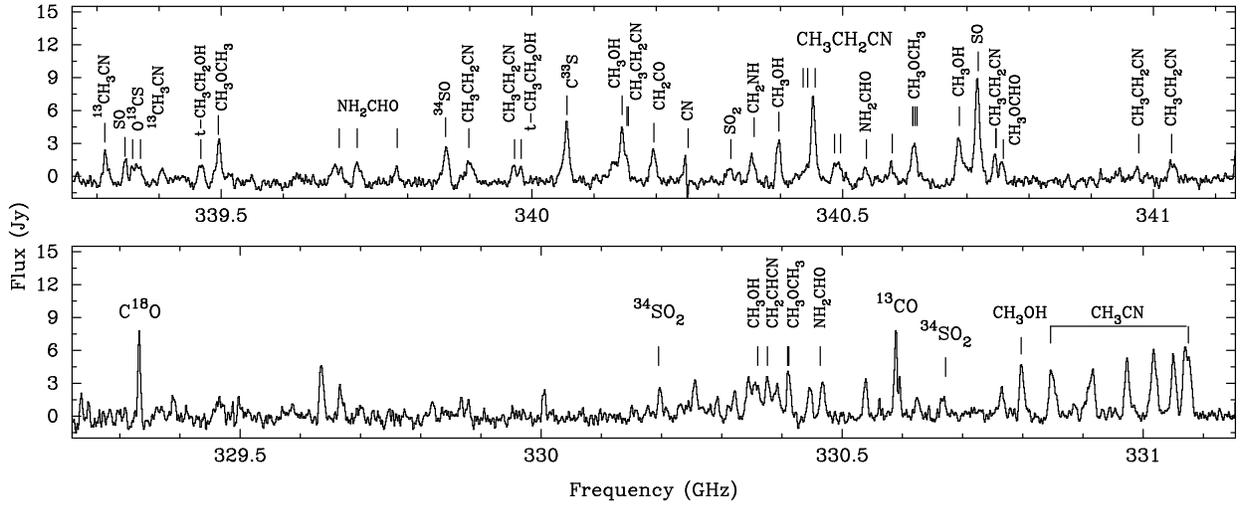}
\caption{\label{spectra-43} 
  SMA spectra toward \irdcfortythree. Both the lower and upper
  sidebands are shown (upper and lower panels respectively). The
  spectra are averaged over a $\sim$ 2\arcsec\, region centered on the
  peak in the high-angular resolution continuum image. Note the
  strong emission from the many complex molecules. Thus, we suggest
  that \irdcfortythree\, is a HMC associated with the
  very earliest stage in the formation of a high-mass protostar. The
  molecular lines marked on this figure were based on the output from
  the Lovas on-line catalog of suggested rest frequencies.}
\end{figure*}

\begin{figure*}[ht]
\centering
\includegraphics[angle=-90,width=0.99\textwidth]{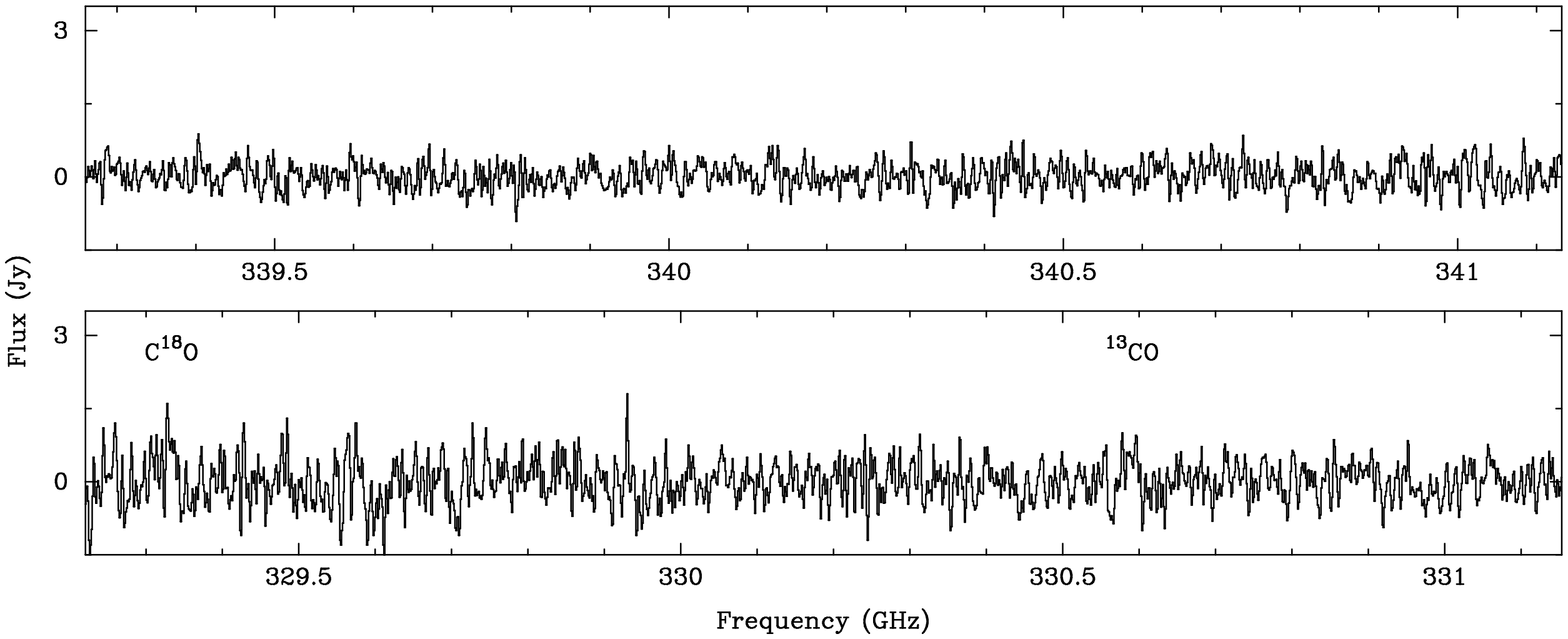}
\caption{\label{spectra-30} 
  SMA spectra toward \irdcthirty. Both the lower and upper sidebands
  are shown (upper and lower panels respectively). The spectra are
  averaged over a $\sim$ 2\arcsec\, region centered on the peak in the
  high-angular resolution continuum image. While \tco\, and \ceo\,
  emission is clearly seen toward this core in the lower-angular
  resolution data (Fig.~\ref{irdc-30}; 14\arcsec\, beam), very little
  is seen in this spectrum. The absence of strong \tco\, or \ceo\,
  emission in the high-angular resolution spectrum may imply that the
  emission seen in the lower-angular resolution spectrum arises from a
  cold extended envelope which is resolved out by the interferometer.}
\end{figure*}

Figures~\ref{spectra-43} and \ref{spectra-30} display the average
spectra within a $\sim$ 2\arcsec\, box centered on the peak in the
high-angular resolution continuum image toward \irdcfortythree\, and
\irdcthirty\, respectively. Toward \irdcfortythree, the spectra show strong emission from
the \tco, \ceo, \so, and \iso\, transitions in addition to numerous
lines from the complex molecules NH$_{2}$CHO, CH$_{2}$NH,
CH$_{3}$OCH$_{3}$, CH$_{3}$CN, and CH$_{3}$CH$_{2}$CN.  Such complex
molecules are typically produced and excited in the immediate
surroundings of a recently formed high-mass protostar as the grain ice
mantles are evaporated (see \citealp{Kurtz00} and references
therein). Moreover, the presence of these species in such a compact
region ($\lesssim$ 0.03 pc) indicates that this is a HMC and, thus, an
early stage in the formation of a high-mass protostar.

\section{Discussion}

\subsection{Gas masses, Jeans masses, Virial masses, and Jeans lengths}

The optically thin sub-millimeter continuum emission can be used to estimate
the gas masses (M$_{G}$) of the condensations via the expression \citep{Hildebrand83}
\[ M_G = \frac{F_{\nu} D^{2}}{\kappa_{\nu} B_{\nu} (T_D)} \]
\noindent where $F_{\nu}$ is the observed integrated 
source flux density, $D$ is the distance, $\kappa_{\nu}$ is the dust
opacity per gram of dust, and $B_{\nu}(T_D)$ is the Planck function at
the dust temperature (T$_{D}$). We estimate $\kappa_{0.87 mm}$ by
scaling the value of $\kappa_{1.3 mm}$ (1.0~cm$^2$ g$^{-1}$;
\citealp{Ossenkopf94}) by $\nu^{\beta}$, where $\beta$, the dust
emissivity index. We assume a gas-to-dust mass ratio of 100. In both
cases we assume gray-body emission but with different dust
temperatures and emissivity indices. Fits to the IR--millimeter
spectral energy distributions (SEDs;
\citealp{Rathborne08}) for each core reveal dust temperatures of 34\,K and 16\,K and
dust emissivity indices of 1.5 and 1.2 for
\irdcfortythree\, and \irdcthirty\, respectively (see Table~\ref{core-properties}). 
We assume these values for $\beta$ in the calculation of the
condensation masses. For \irdcfortythree\, we assume a T$_{D}$ of
100\,K because this is a typical temperature for HMCs
\citep{Kurtz00}.

An assumption of the dust temperature for the condensations within
\irdcthirty\, is more difficult. Although fits to the core SED reveal 
a T$_{D}$ of 16\,K, it is possible that the condensations may be small
volumes of slightly warmer gas within this extended, cold envelope.
Because strong emission at sub-millimeter wavelengths can trace either
temperature or density enhancements ideally one needs high-angular
resolution molecular line observations to directly measure the
temperatures (e.g., \ammonia). The non-detection of any molecular
lines in the high-angular resolution spectrum suggests that the
temperature of the gas is quite cold.  Thus, to calculate the masses
for the condensations within \irdcthirty\, we assume a temperature of
16\,K. Because the temperature of the condensations may be higher than
the temperature of the bulk of the gas within the core, the derived
condensation masses may, therefore, be an upper limit to the true
mass. For example, if the condensations have a temperature of 30\,K
rather than 16\,K, the derived masses will decrease by a factor of
$\sim$ 2.

At high-angular resolution, the derived mass for \irdcfortythree\, is
29\,\Msun, while for \irdcthirty\, the condensation masses are
9--21\,\Msun\, (Table~\ref{condensations}).  The lower-angular
resolution single dish millimeter continuum images reveal these cores
to have masses of 430 and 1,130\,\Msun\, respectively\footnote{These
masses differ slightly from those quoted in \cite{Rathborne06}
because of their assumption of a single dust temperature (15 K) and
emissivity index (2.0) for all their cores.}. Thus, we find that only
a small fraction ($\sim$ 4\%) of the mass detected in the
lower-angular resolution millimeter continuum data is recovered by the
interferometer. Thus, it is likely that the emission traced by the
interferometer corresponds to well-defined small, dense volumes of gas
within a larger, more diffuse core.

To determine the likelihood that the cores and condensations will
fragment further, we calculate both their Jeans (M$_J$) and Virial (M$_V$) masses. 
These masses were calculated using the expressions

\[ M_J \approx  1.1 \,M_{\odot}  \left [ \frac{T_D}{ 10 {\mathrm {\,K}}} \right]^{3/2} \left [ \frac  {\rho} {10^{-19} {\mathrm {\,g\, cm^{-3}}}} \right]^{-1/2} \]

\[M_{\rm V} = 1.145\times10^{3}\,\,\Delta V ^{2}\, D\, \sqrt{\Omega} \hspace{0.8cm} M_{\odot}\]

\noindent where $T_{D}$ is the dust temperature, $\rho$ is the mass density
\citep{Zinnecker07}, $\Delta$ V is the measured line width (\kms), $D$ is 
the distance (kpc), and $\Omega$ is the area on the sky (deg$^{2}$).
The mass density was estimated using the dust masses and by assuming
the volume is equivalent to a sphere of radius $R$.

The Jeans masses for the cores and condensations were calculated
assuming the dust temperatures listed in Tables~\ref{core-properties} and
\ref{condensations} and assuming that the gas and dust temperatures
are identical.  One must exercise caution, however, in applying a
temperature of 34 K for the core \irdcfortythree\, as it is associated
with a luminous 24\,\um\, source which may be heating the gas to well
above 34 K. For a gas temperature of 100~K, M$_{J}$ will be higher by
a factor of $\sim$10. While the HMC's temperature may be closer to
100\,K and correspond to a small volume in the core, the bulk of gas
is likely to be colder. If the actual gas temperature is higher by a
factor of two than the assumed value, then M$_{J}$ will increase by a
factor of $\sim$4.

In the case where thermal processes dominate, the ratio of the gas
mass to Jeans mass (M$_G$/M$_J$) estimates the degree to which a gas
structure is likely to gravitationally collapse into thermally
fragmented substructures. In contrast, if turbulence dominates, then
the Jeans mass is no longer applicable and one must consider the
Virial mass. In this case, the ratio of the gas mass to Virial mass
(M$_{G}$/M$_{V}$) is a better indicator of further fragmentation.  In
either case, however, a ratio greater than unity indicates that the
gas is unstable and will further fragment.

Because the core line profiles in Figures~\ref{irdc-43} and
\ref{irdc-30} clearly show that the lines are much broader than what is
expected from thermal broadening alone, it is likely that turbulence
may significantly contribute to the pressure support within both
cores.  While we list the ratio M$_G$/M$_J$ for the cores and
condensations in Tables~\ref{core-properties} and \ref{condensations}
it is unlikely that this ratio is physically significant.  Instead, we
use the ratio M$_{G}$/M$_{V}$ to estimate the degree to which a gas
structure is likely to gravitationally collapse into fragmented
substructures.

To calculate \Mvirial\, for the cores we use the line
widths measured from the low-angular resolution optically thin \ceo\,
emission (see Table~\ref{lines}).  For the condensations, we assume
the same line width as measured toward the larger core.  Because
high-angular resolution NH$_{3}$ observations toward a nearby IRDC
show typical line widths of $\sim$ 1.2\,\kms\, \citep{Wang06} on
scales of the condensations, using the line width from the larger core
may result in an overestimation of the calculated \Mvirial\, for the
condensations.

We find that M$_{G}$/M$_{V}$ is 2.8 and 2.3 for the cores
\irdcfortythree\, and \irdcthirty\, respectively (Table~\ref{core-properties}).
For the condensations, we find that M$_{G}$/M$_{V}$ is $\lesssim$ 1
(Table~\ref{condensations}).  Since M$_{G}$/M$_{V}$ is greater than
unity for the cores, they are likely to fragment, but since
M$_{G}$/M$_{V}$ is less than unity for the individual condensations,
they are stable against further fragmentation and thus will probably
give rise to individual stars.



The Jeans length for the cores and condensations was calculated
using the expression
\[ \lambda_J = \sqrt{\frac{\pi k T_D}{G m_{p} \rho}} \]
where $k$ is the Boltzmann constant, $T_{D}$ is the dust temperature,
$G$ is the gravitational constant, $m_{p}$ is the mass of a proton,
and $\rho$ is the mass density.

We find that the Jeans length, defined as the minimum length scale
(radius) for gravitational fragmentation to occur, is smaller
($\sim$0.04\,pc) for \irdcfortythree\, than that measured toward
\irdcthirty\, ($\sim$0.16\,pc; Table~\ref{core-properties}). The 
separation of the condensations within \irdcthirty\, ($\sim$0.2\,pc)
is comparable to the Jeans length of this core.  Moreover, on the
smallest scales of the individual condensations, the Jeans lengths are
comparable to the measured radii of the condensations
(Table~\ref{condensations}). These results support the idea that
gravitational collapse is occurring within the IRDCs and that the
fragmentation process has stopped at the smallest protostellar scales.

\subsection{\irdcfortythree: a high-mass protostar with a rotating envelope?}

The core \irdcfortythree\, shows many indicators of high-mass star
formation, including a bright 24\,\um\, point source, broad molecular
line emission, shocked gas, water maser emission, and molecular outflows
\citep{Garay04,Shepherd04,Rathborne05,Rathborne06,Wang06}.
\irdcfortythree\, also shows weak radio continuum emission consistent 
with a deeply embedded B2 protostar \citep{Shepherd04}. Fits to its
IR--millimeter SED reveal it contains an object with a bolometric luminosity of
$\sim$ 10$^{4.3}$\,\Lsun, consistent with a B0 protostar
\citep{Rathborne08}.  Because its high-angular resolution
sub-millimeter continuum flux is dominated by a single, compact,
massive emission feature, it is likely that the luminosity is
dominated by a single high-mass protostar.

\begin{figure}[ht]
\centering
\includegraphics[angle=-90,width=0.45\textwidth]{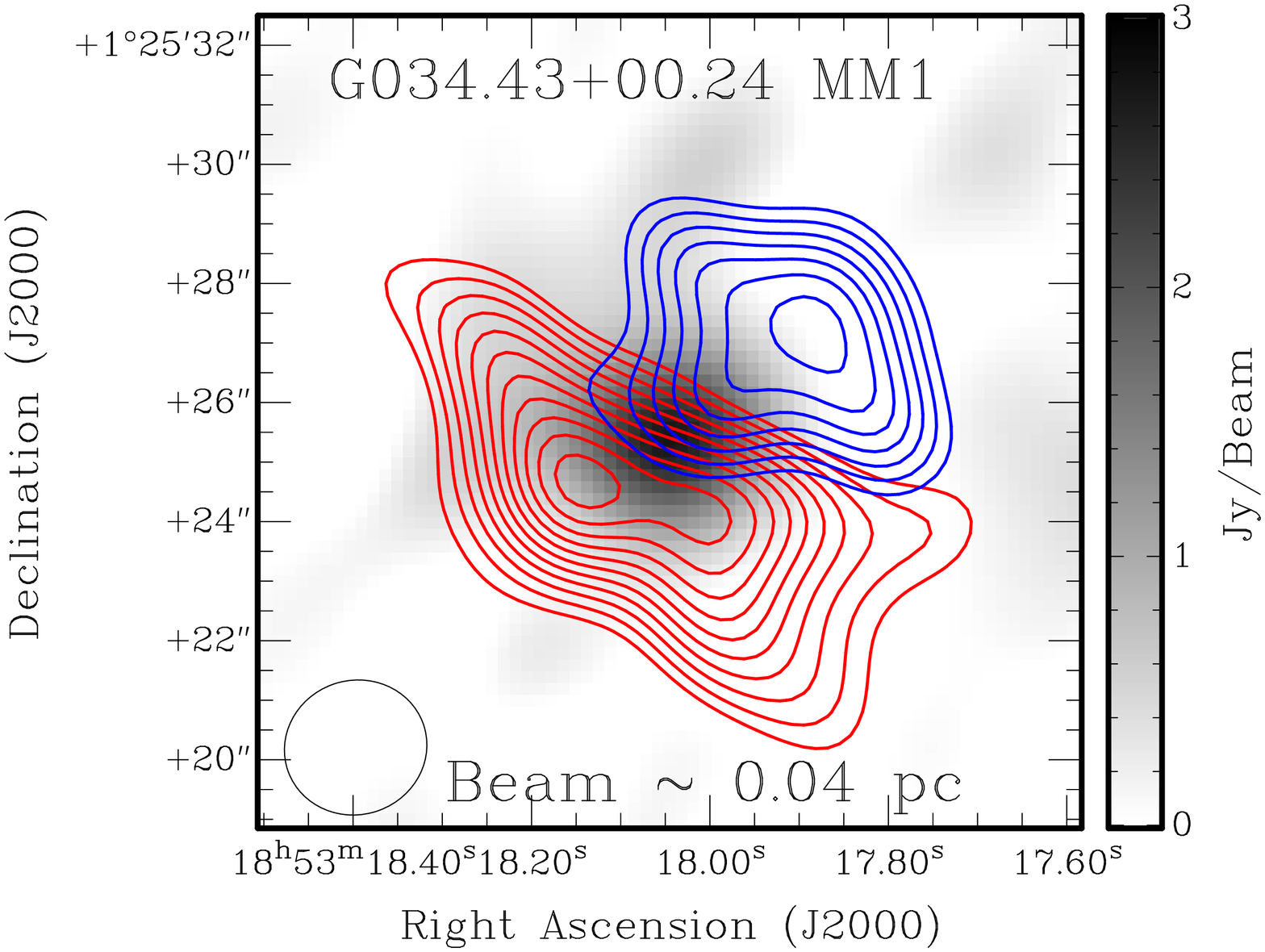}
\caption{\label{kinematics} SMA \chtchtcn\, integrated intensity image 
      in gray scale tracing the unresolved HMC within
      \irdcfortythree. Contours show the red-shifted (\vlsr =
      61\,\kms) and blue-shifted (\vlsr= 55\,\kms) \tco\, emission
      surrounding the unresolved, bright HMC (emission is from a
      single channel, with the same contour levels for the red and
      blue components; 1.7 to 7.5 in steps of 0.5 Jy
      beam$^{-1}$). This emission appears to be tracing an extended
      rotating envelope.}
\end{figure}

Additional clues to the nature of this high-mass star-forming HMC 
can be obtained by examining the morphologies of its molecular line
emission. For example, Figure~\ref{kinematics} shows a distinct
morphological difference between the \chtchtcn\, and
\tcont\, emission. The `hot core
lines', in this case traced by the \chtchtcn\, emission, are all
unresolved at $\sim$0.03\,pc resolution, while the \tcont\, emission
is easily resolved into a $\sim$ 0.12 pc diameter extended region.
Because the hot cores lines are excited in the immediate vicinity of a
high-mass protostar, it is likely, therefore, that the unresolved
\chtchtcn\, emission is tracing the heated, compact ($\lesssim$0.03
pc, $\lesssim$ 6,500 AU) regions immediately surrounding the central
protostar. On the other hand, the extended \tcont\, emission is
tracing the larger, less dense circumstellar region of the core.

This extended \tcont\, circumstellar emission also reveals a clear
velocity gradient with red-shifted emission to the southeast and
blue-shifted emission to the northwest of the central protostar
(Fig.~\ref{kinematics}).  Such a velocity gradient might arise from
either an outflow or a rotating envelope. Observational evidence for
accretion disks around high-mass protostars exists for only a few
objects (see \citealp{Cesaroni07} and references therein). The
widespread existence of large ($>$ 10,000 AU) disk-like structures in
young HMCs would support the idea that high-mass stars form via
monolithic collapse (e.g., \citealp{McKee03}) rather than via
competitive accretion (e.g. \citealp{Bonnell02}).  Because HMCs in IRDCs
are probably very young, they are excellent candidates to search for
high-mass accretion disks given their short lifetimes.

In recent OVRO observations of CO (1-0) emission, \cite{Shepherd07}
identify two outflows that may originate within the core
\irdcfortythree. The dominant red-shifted lobe of the main outflow
(their G34.4:A) appears to extend to the southwest, with little
blue-shifted emission found to the northeast. Moreover, high-angular
resolution SiO (1--0) imaging of this core with the VLA (Q. Zhang
priv. comm.) confirms the presence and orientation of the CO outflow.
This outflow appears to be orientated perpendicular to the extended
\tco\, structure found here.  Thus, the orientation of the outflow 
is consistent with this extended \tco\, structure being a rotating
envelope surrounding the central high-mass protostar.

\begin{figure}[ht]
\centering
\includegraphics[width=0.45\textwidth]{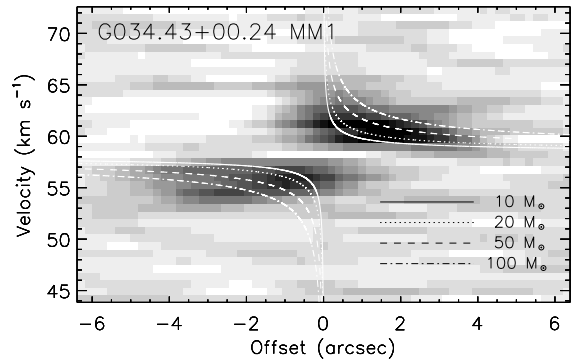}
\caption{\label{lv} SMA \tco\, position-velocity diagram for \irdcfortythree. 
    The slice is taken perpendicular to the axis of the rotating
    envelope and shows the two components. Overlaid on this image are
    models for Keplerian rotation with central masses of 10, 20, 50,
    and 100\,\Msun. The emission from \irdcfortythree\, is consistent
    with Keplerian rotation around a $\sim$ 10--50\,\Msun\, source.}
\end{figure}

The position--velocity diagram of the circumstellar \tco\, emission
toward \irdcfortythree\, is consistent with Keplerian rotation around
a $Msin(i)$ $\sim$ 10--50\,\Msun\, protostar (Fig.~\ref{lv}). Thus, we
see tentative evidence for a rotating circumstellar envelope
surrounding this protostar. The derived mass for the central protostar
is consistent with the luminosity and spectral type determined from
the core's SED (10$^{4.3}$\,\Lsun; B0).  Higher angular and spectral
resolution molecular line observations are required to better resolve
the velocity field in order to confirm or refute the presence of an
accretion disk.

\subsection{\irdcthirty: an early phase of high-mass star-formation?}

In contrast to the bright, rich molecular line spectrum toward
\irdcfortythree, the lack of strong molecular line emission toward
\irdcthirty\, suggests that it is either colder, more extended, or
both. While \tco\, and \ceo\, emission is clearly seen toward this
core in the lower-angular resolution data (Fig.~\ref{irdc-30};
14\arcsec\, beam), very little is seen in the high-angular resolution
spectrum (Fig.~\ref{spectra-30}; upper limit of $\sim$ 2.9
K). Moreover, in the 14\arcsec\, single dish spectra, \irdcthirty\,
shows only faint HCN (4--3) emission but does show self-absorbed CS
(3--2), saturated or self-absorbed \nthp, broad SiO (2--1) and \htcop,
and an asymmetry in the \hcop\, emission. The lack of emission from
the high excitation HCN transition and the saturated \nthpnt\, profile
suggests that the molecular gas is cold with a very high optical
depth. Because SiO emission typically traces outflow activity caused
by shocked gas, it is usually detected only in star-forming cores.
Thus, the detection of broad line wings in the SiO, \hcopnt\, and
\htcopnt\, spectra suggests that this core may already harbor some
star formation activity.

The absence of strong \tco\, or \ceo\, emission in the high-angular
resolution spectrum may imply that the emission seen in the
lower-angular resolution spectrum arises from a cold extended envelope
which is resolved out by the interferometer. Using the limit of $\sim$
2.9 K for the brightness sensitivity of the SMA spectrum, we estimate
from the single dish spectra that the sizes of the \tco\, and
\ceo\, emitting regions must be $>$ 16\arcsec\, ($>$0.4 pc). However, 
converting the derived mass to a \tcont\, column density, we predict
that the individual condensations have sufficient column densities
($\sim$ 10$^{23}$\,\cms) to be detected in \tco\, emission in the
current observations (peak T $\sim$ 10 K and $\Delta$V=5\,\kms).  It
might be, therefore, that the \tcont\, is depleted toward these
condensation. It appears that \irdcthirty\, does not contain any
volume of heated gas and likely is a massive, cold core. If a
protostar does exist within \irdcthirty\, then it has not yet heated
its surrounding gas.  Moreover, if it does contain a circumstellar
disk, then it would be much bigger and colder compared to the disk
surrounding
\irdcfortythree.  

Given the core's mass is $\sim$ 1000\,\Msun, and assuming a
star-formation efficiency of 30\%, we estimate that this core should
give rise to a cluster of 300\,\Msun.  Using the expression
$m_{max}$ = 1.2 $M_{cluster}^{0.45}$ \citep{Larson03}, which relates
the maximum stellar mass in a cluster ($m_{max}$) to the mass of the
cluster ($M_{cluster}$), we estimate $m_{max}$ to be $\sim$
16\,\Msun. Thus, given its mass and a typical initial mass function
(IMF), we would expect this core to give rise to $\sim$ 4 high-mass
($>$ 8\,\Msun) stars. Because it contains compact condensations, some
evidence for star-formation activity but no obvious heating, we
suggest that \irdcthirty\, may be an example of a very early stage in
high-mass star-formation. The measured bolometric luminosity
($\sim$10$^{2}$\,\Lsun) may arise from the accretion of material
onto the condensations. Given that the most massive condensation may
further fragment it is possible that these cold condensations will
continue to accrete material and give rise to a cluster of several
high-mass protostars.

It is interesting to compare our observations with recent models that
suggest that the fragmentation of a core can be suppressed by the
increase in its temperature due to accretion onto lower-mass
protostars \citep{Krumholz06,Krumholz08}. In this scenario, a core's
initial temperature and Jeans mass predicts it will fragment into many
low-mass condensations. However, as these condensations form and begin
to accrete, the temperature increases which, in turn, increases the
overall Jeans mass of the core. An increase in temperature to $\sim$
100~K is thought to be sufficient to halt the fragmentation
process. To form high-mass cores and avoid further fragmentation,
\cite{Krumholz08} find that the cloud's critical column density,
$\Sigma$, should be $\gtrsim$ 1\,g\,cm$^{-2}$ and its
luminosity-to-mass ratio, L/M, $\gtrsim$ 10
\Lsun/\Msun. For the core \irdcthirty\, we find that $\Sigma$ is $\sim$
0.4\,g\,cm$^{-2}$ and L/M is $\sim$ 0.3\,\Lsun/\Msun\,
(Table~\ref{core-properties}).  These rough estimates are smaller than
the predicted thresholds. Thus, fragmentation and the formation of
lower-mass protostars is still possible within \irdcthirty. This is in
contrast to \irdcfortythree\, which has a measured $\Sigma$ and L/M
much greater than these thresholds implying that it will give rise to
a high-mass star. In this case, however, its high M$_{G}$ to M$_{J}$ ratio (for
T$_{D}$ = 100 K) suggests that the increase in gas temperature alone
may not be enough to halt thermal fragmentation.

The ongoing fragmentation within \irdcthirty\, is consistent with our
detection of multiple condensations toward its center. If this model
is correct, then it appears that only the condensation A has
sufficient column density to form a high-mass star, the other
condensations will most likely only form low-mass stars
(Table~\ref{condensations}). The derived temperature for the dust
within this core is well below 100 K, suggesting that material has not
been heated sufficiently to halt the fragmentation. Due to its high
mass, this represents an excellent core in which to test these
theoretical predictions for high-mass star-formation.

\section{Conclusions}

Using the SMA we have obtained high-angular resolution sub-millimeter
continuum images and molecular line spectra toward two cores within
IRDCs. In the low-angular resolution data these cores have sizes R
$<$0.8\,pc, gas masses M $>$400\,\Msun, and bolometric luminosities
L~$\gtrsim$10$^{2}$\,\Lsun\, \citep{Rathborne06,Rathborne08}.  Despite
their similar sizes, masses, and bolometric luminosities the two cores
show remarkable differences in the high-angular resolution data. The
sub-millimeter continuum images reveal that
\irdcfortythree\,remains unresolved and has a size of $\sim$ 0.03 pc and mass of 
29\,\Msun. In contrast, \irdcthirty\, contains at least three compact
($\sim$ 0.06 pc), massive (9--21\,\Msun) condensations.

Moreover, the molecular line spectra reveal that \irdcfortythree\,
emits strong lines from many complex molecules, but the core
\irdcthirty\, does not. This suggests that \irdcfortythree\, is hot, compact, 
and dense, while \irdcthirty\, is colder and possibly more extended.
We speculate, therefore, that \irdcfortythree\, is a HMC in an early
stage in the formation of a high-mass protostar, while \irdcthirty\,
may represent an even earlier phase in the high-mass star-formation
process.  Because \irdcthirty\, shows some evidence for active star
formation in the low-angular resolution data (e.g., shocked gas, high
velocity line profiles, and a high bolometric luminosity), we cannot
rule out the possibility that the condensations within this core may
have already begun to accrete. However, the low temperatures and
absence of bright 24\,\um\, point sources or strong molecular line
emission from high excitation transitions suggests that there is
little localized heating. Because of its size and mass, \irdcthirty\,
is likely to give rise to at least one high-mass star. We speculate
that the detected condensations may continue to accrete material from
their surroundings to eventually form high-mass stars. Since the
condensations account for only $\sim$4\% of the total mass, there is
an ample amount of supply of material to accrete.

The properties of these cores support the idea that high-mass star and
cluster formation likely occurs in IRDCs. High-angular sub-millimeter
continuum images and molecular line spectra toward more cores within
IRDCs are needed to unambiguously establish the evolutionary path from
a high-mass starless core to a high-mass protostar.

\acknowledgments
The authors gratefully acknowledge funding support through NASA grant
NNG04GGC92G. This work is based in part on observations made with the
{\it Spitzer Space Telescope}, which is operated by the Jet Propulsion
Laboratory, California Institute of Technology under NASA contract
1407. Support for this work was provided by NASA through contract
1267945 issued by JPL/Caltech. The JCMT is operated by JAC, Hilo, on
behalf of the parent organizations of the Particle Physics and
Astronomy Research Council in the UK, the National Research Council in
Canada, and the Scientific Research Organization of the
Netherlands. IRAM is supported by INSU/CNRS (France), MPG (Germany),
and IGN (Spain). We would also like to thank Steven Longmore and Mark
Krumholz for useful discussions. The authors would also like to thank
the anonymous referee for comments that greatly improved the paper.

{\it Facilities:} \facility{SMA, IRAM, JCMT, Spitzer}.


\clearpage

\begin{table*}
{\scriptsize
\centering
\caption{\label{core-properties}Core properties}
\begin{tabular}{ccrcccccccccccc}
\tableline
\tableline
Core & \multicolumn{2}{c}{Coordinates}        & \vlsr  & D         & R        & T$_{D}$ & $\beta$   & L           &  M$_{G}$               & $\frac{M_{G}}{M_{J}}$   & $\frac{M_{G}}{M_{V}}$  & $\lambda_{J}$  & $\Sigma$ & $\frac{L}{M}$\\
     & RA       & \multicolumn{1}{c}{Dec}     &        &           &          &         &           &             &                        &                         &                        &                &          &\\          
     & (J2000)  & \multicolumn{1}{c}{(J2000)} & (\kms) & (kpc)     & (pc)     & (K)     &           & (\Lsun)     &(\Msun)                 &                         &                        & (pc)           &          &\\
\tableline 									  		                      	                                	                                                                   
\irdcfortythree & 18:53:18.0 &    01:25:24   & 57     &  3.7      & 0.10     &  34     & 1.5        &   19,600    & 430                    &  550                    &  2.8                   & 0.04           &  3.2     & 46\\      
\irdcthirty     & 18:44:18.0 & $-$03:59:34   & 87     &  5.7      & 0.42     &  16     & 1.2        &   370       & 1130                   &  800                    &  2.3                   & 0.16           &  0.4     & 0.3\\           
References      &            &               & 1      & 1         &  2       & 3       &  3         & 3           & 3                      &  4                      &  4                     &  4             &  4       & 4 \\            
\tableline
\end{tabular}
\tablerefs{(1) \cite{Simon-msxgrs}; (2) \cite{Rathborne06}; (3) \cite{Rathborne08}; (4) this work.}
}
\end{table*}

\begin{table*}
{\scriptsize
\centering
\caption{\label{lines}Summary of the low-angular resolution molecular line spectra}
\begin{tabular}{lccccccccc}
\tableline
\tableline
Molecular line & \multicolumn{4}{c}{\irdcfortythree} && \multicolumn{4}{c}{\irdcthirty} \\
\cline{2-5} \cline{7-10}
                & \tastar & \vlsr & $\Delta$V & I\tablenotemark{a} && \tastar & \vlsr & $\Delta$V & I\tablenotemark{a}\\
                & (K) & (\kms) & (\kms) & (K\,\kms) && (K) & (\kms) & (\kms) & (K\,\kms) \\
\tableline 									  				                                              
\ceo            &  2.9  &  57.8 &   3.7 & 11.1 &&  1.2 &  87.0 &  3.2 &  4.8 \\
\tco            &  8.2  &  57.6 &   5.1 & 44.8 &&  2.4 &  86.2 &  5.8 & 14.8\\
HCN (4--3)      &  3.4  &  57.7 &  12.5 & 45.8 &&  1.2 &  86.0 &  7.6 &  1.1\\
CS (3--2)       &  4.1  &  57.9 &   6.8 & 31.7 &&  0.5 &  86.1 &  7.4 &  4.2\\
\nthp           &  1.0\tablenotemark{b}  &  57.8 &   2.6 & 57.1&&  0.2\tablenotemark{b}  &  86.8 &  3.6 & 16.4\\
SiO (2--1)      &  0.8  &  57.9 &  10.6 & 9.3  &&  0.3 &  86.4 &  9.0 &  3.2\\
\htcop          &  0.8  &  57.9 &  10.5 & 9.0  &&  0.3 &  86.4 &  9.1 &  3.3\\
\hcop           &  2.5  &  60.2 &   3.7 & 13.4 &&  1.1 &  87.5 &  9.2 & 11.2\\
\tableline
\end{tabular}
\tablenotetext{a}{The integrated intensity quoted here is not the area under the Gaussian, but instead is calculated over a velocity range to include all the emission.}
\tablenotetext{b}{This is $\tau$\tastar\, for the main hyperfine components.}
}

\end{table*}

\begin{table*}
{\scriptsize
\centering
\caption{\label{condensations}Condensation properties}
\begin{tabular}{lcrccccccccc}
\tableline \tableline
\multicolumn{1}{c}{Condensation} &\multicolumn{2}{c}{Coordinates}          & \multicolumn{1}{c}{Peak} & Deconvoled     & T$_{D}$ & $\beta$   &  \multicolumn{1}{c}{M$_{G}$}   & $\frac{M_{G}}{M_{J}}$  & $\frac{M_{G}}{M_{V}}$& $\lambda_{J}$   & $\Sigma$\\
                                 &  RA & \multicolumn{1}{c}{Dec}           & \multicolumn{1}{c}{flux} & diameter       &         &           & 			              &                        &                      &   		         &         \\        
                                 & (J2000) & \multicolumn{1}{c}{(J2000)}   & \multicolumn{1}{c}{(mJy)}& (pc)           & (K)     &           & \multicolumn{1}{c}{(\Msun)}    &                        &                      & (pc)                 &         \\          
\tableline													           	 	    							                               	                  
\irdcfortythree                  & 18:53:18.03 &  1:25:25.58               &   2760                   &  0.03         &  100    & 1.5       & 29                              &  30                    &  1.2                 & 0.02    	         &  7.7    \\         
\irdcthirty~A                    & 18:44:18.08 &  $-$3:59:34.33            &   73                     &  0.05         &  16     & 1.2       & 21                              &  160	               &  0.8	               & 0.01  	         &  2.7	\\          
\irdcthirty~B                    & 18:44:17.95 &  $-$3:59:31.27            &   33                     &  0.10         &  16     & 1.2       & 9	    		              &  10 	               &  0.2 	               & 0.07  	         &  0.2	\\          
\irdcthirty~C                    & 18:44:17.93 &  $-$3:59:35.42            &   31                     &  0.06         &  16     & 1.2       & 9			              &  30  	               &  0.3 	               & 0.03  	         &  0.8    \\         
\tableline
\end{tabular}
}
\end{table*}

\clearpage

\end{document}